\documentclass[aps,twocolumn,amsmath,letterpaper]{revtex4}
\usepackage{amssymb}
\usepackage{graphicx}
\usepackage{array}
\usepackage{hhline}
\usepackage{longtable}

\begin{document}

    \title{A Comparative Study of Interdisciplinarity in Sciences\\
in Brazil, South Korea, Turkey, and USA}

\author{Nazl\i~Yurdakul$^{1}$ and A. Nihat Berker$^{2,3}$}
\affiliation{$^1$Robert College, Arnavutk\"oy 34345, Istanbul,
Turkey} \affiliation{$^2$Faculty of Engineering and Natural
Sciences, Sabanc\i~University, Tuzla 34956, Istanbul, Turkey}
\affiliation{$^3$Department of Physics, Massachusetts Institute of
Technology, Cambridge, Massachusetts 02139, U.S.A.}

\begin{figure*}[] \centering
\includegraphics*[scale=0.3]{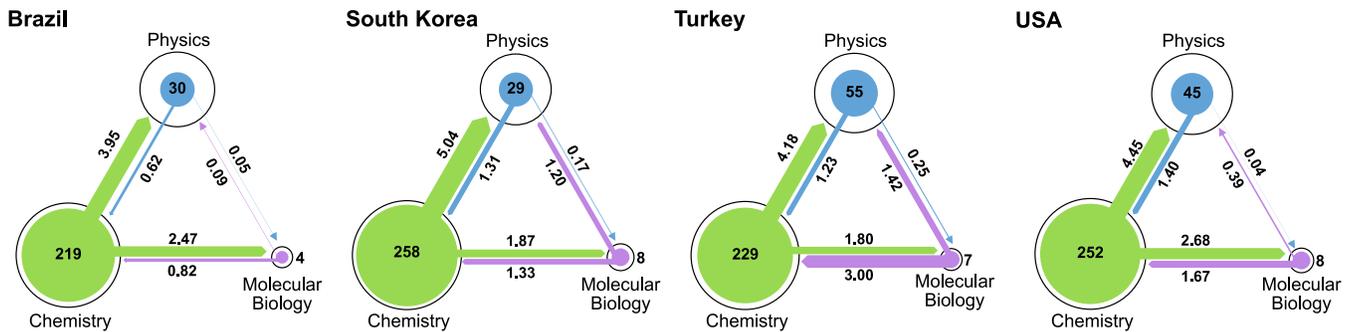}
\caption{Interdisciplinary citations given in 2013, as described in
the text, between chemistry, molecular biology, and physics, in
Brazil, South Korea, Turkey, and USA. The direction of each arrow is
from the field giving citations towards the field being cited. The
width of each arrow is proportional to the average number of such
citations per publication, also written next to the arrow. In a
given field, approximately the same number of publications is used
for each country. Thus, the area inside the drawn circle is
proportional to the total number of publications in the pool. For
each country and each field, the area of the colored circle is
proportional to the total number of papers giving such cross-science
citations, also written inside or next to the colored circle.}
\end{figure*}

\begin{abstract}

A comparative study is done of interdisciplinary citations in 2013
between physics, chemistry, and molecular biology, in Brazil, South
Korea, Turkey, and USA. Several surprising conclusions emerge from
our tabular and graphical analysis: The cross-science citation rates
are in general strikingly similar, between Brazil, South Korea,
Turkey, and USA. One apparent exception is the comparatively more
tenuous relation between molecular biology and physics in Brazil and
USA. Other slight exceptions are the higher amount of citing of
physicists by chemists in South Korea, of chemists by molecular
biologists in Turkey, and of molecular biologists by chemists in
Brazil and USA. Chemists are, by a sizable margin, the most
cross-science citing scientists in this group of three sciences.
Physicist are, again by a sizable margin, the least cross-science
citing scientists in this group of three sciences. In all four
countries, the strongest cross-science citation is from chemistry to
physics and the weakest cross-science citation is from physics to
molecular biology. Our findings are consistent with a V-shaped
backbone connectivity, as opposed to a $\Delta$ connectivity, as
also found in a previous study of earlier citation years.

\end{abstract}

    \maketitle
    \def\s{\rule{0in}{0.28in}}
    \setlength{\LTcapwidth}{\columnwidth}

\section{Introduction}

While interdisciplinarity is currently much vaunted as the
scientific mode of operation, intense specialization in any one
field or, in fact, topic may run counter to cross-disciplinary
efforts.  Another characteristic of current science is the
burgeoning of a multicontinental multicenter research environment,
which brings the question of whether different regional, historical
and current, academic traditions affect the conduct of scientific
research. We have investigated simultaneously both of these issues,
by conducting a comparative study between the Brazil, South Korea,
Turkey, and USA, as to the cross-referencing between published
research papers in chemistry, molecular biology, and physics.  Our
interdisciplinary and academic intercultural findings, based on
collected data, are surprising on both of the mentioned issues.

Our study involves cross-disciplinary citations between fields A and
B, where A and B are chemistry, molecular biology, and physics, a
priori deemed derivatively connected basic sciences, in articles
published in a set of major journals (Tables I-III) in each field in
the year 2013. The study is repeated for Brazil, South Korea, USA,
and Turkey. These countries were chosen because of the dominance in
scientific research of the USA, and the rapid development of the
transcontinentally and mutually distant Brazil, South Korea, and
Turkey. Our study was inspired by Ref.\cite{RosvallBergstrom}, where
the cross-citation network between fields is studied for earlier
years, without distinguishing with respect to country. Similar
studies have been made for the citation network between different
journals in the same field \cite{RosvallBergstrom2} and on the
relevance of cross-science citations \cite{RosvallBergstrom3}.
Detailed intercultural comparative studies are in
Refs.\cite{Usdiken, Schott1, Schott2,Hwang}.

\section{Methodology}

In our study, 67, 33, 22 journals (Tables I-III), respectively in
chemistry, molecular biology, physics, were used. Of these, 46, 8,
17 journals (emphasized in Tables I-III) were searched for
cross-science citing publications as described below and yielded
958, 26, 159 cross-science citing publications, given to 116, 199,
161 journals. Thus, 7696, 138, 756 cross-science citations were
given from respectively chemistry, molecular biology, physics, by
authors with institutional addresses in Brazil, Turkey, South Korea,
or USA. These cross-science citations were given as 777, 2649, 5164
to respectively chemistry, molecular biology, physics. In these,
publications with author addresses from more than one of our studied
countries were not included. Thus, a total of 8590 cross-science
citations entered our study.

In order to effectively compare the citation practices from each
country, the pool of sample publications in each science must be as
similar as possible between the countries.  The number of
publications by Brazilian, South Korean, and USA scientists in 2013
exceeds those by Turkish scientists in most, but not all, of the
selected chemistry, molecular biology, and physics journals (Tables
I-III). Therefore, the sample size of Brazilian, South Korean, and
USA papers was equalized to the number of Turkish papers published
in 2013: The Brazilian, South Korean, and USA publications in each
journal were ordered chronologically.  Then, in each journal, the
used pool of publications was chronologically expanded equally both
ways starting from the median publication until the number of
publications was equalized to that of Turkish publications in the
same journal in 2013.  For example, there are 17 papers published by
Turkish physicists in the Physical Review A in 2013. Thus, the
chronologically median publications in Physical Review A in 2013 by
Brazilian, South Korean, and USA physicists were found and the pool
was expanded equally in both chronological direction until there
were 17 papers in the pool from each country. In several cases, the
number of Turkish publications in a given journal exceeded the
number of Brazilian, South Korean, or USA publications. In these
instances, the pool of Turkish publications was not decreased and
all of the Brazilian, South Korean, or USA publications were
included.

The same pool of publications, for each country and each science,
was used for determining the citation flow from this science to each
of the two other sciences. For instance, there were 158 physics
publications by Turkish authors in the selected journals. This same
set of 158 papers was used to determine the average number, per
publication, of citations to chemistry and to molecular biology. The
standard deviation was also determined. When calculating the average
and the standard deviation, citations to all publications in the
other science are of course included, regardless of the country of
the publication receiving the citation.   The results are given in
Fig. 1 and Tables IV-VI.

\section{Results and Discussion}

In Fig. 1, for each country and each science, the area of the
colored circles is proportional to the total number of publications
giving cross-science citations to the two other sciences, also given
numerically inside or next to the colored circles. The area inside
the drawn circles is proportional to the total number of
publications considered. Therefore, as explained above, for each
field the latter areas are similar, but not strictly equal, between
the countries. The widths of the arrows are in turn proportional to
the average number of citations, per publication, from the field
they originate to the field they are pointing. The corresponding
numerical data are given next to the arrows and in Tables IV-V.

Several surprising conclusions emerge from these data: 1) The
cross-science citation rates are in general strikingly similar,
between Brazil, South Korea, Turkey, and USA. Thus, the common
problems, methodology, instant communications, and personal mobility
in a given science appears to have transcended geographically widely
separated regional cultures. 2) One apparent exception to the above
is the comparatively more tenuous relation between molecular biology
and physics in Brazil and USA. Other slight exceptions are the
higher amount of citing of physicists by chemists in South Korea, of
chemists by molecular biologists in Turkey, and of molecular
biologists by chemists in Brazil and USA. From both items here, it
is seen that Brazil and USA are following a similar (Western
Hemisphere) track. 3) Chemists are, by a sizable margin, the most
cross-science citing scientists in this group of three sciences.
Physicist, although reputed to be more generalists, are, again by a
sizable margin, the least cross-science citing scientists in this
group of three sciences. (Fig.1 and Table VI) 4) In all four
countries, the strongest cross-science citation is from chemistry to
physics and the weakest cross-science citation is from physics to
molecular biology. 5) Our findings are consistent with a V-shaped
backbone connectivity, as opposed to a $\Delta$ connectivity,
consistently with what was found for earlier citation years in
Ref.\cite{RosvallBergstrom}.

\begin{acknowledgments}
We are grateful to Behl\"{u}l \"{U}sdiken for a careful reading of
our manuscript and many useful remarks. We thank Asuman Aky\"{u}z,
Tolga \c{C}a\u{g}lar, Bedia Erim, and Zehra Sayers for advice and
assistance. Support by the Academy of Sciences of Turkey (T\"UBA) is
gratefully acknowledged.
\end{acknowledgments}

\begin{table*}[h!]
\begin{tabular}{! \vrule c c c  ! \vrule }
\hline
&Analytical Chemistry&\\
\hline
\emph{\textbf{Analyst}} &\vline &\emph{\textbf{J. American Society for Mass Spectrometry}} \\
\emph{\textbf{Analytica Chimica Acta}} &\vline &\emph{\textbf{Journal of Chromatography A}}  \\
Analytical and Bioanalytical Chemistry &\vline &\emph{\textbf{Sensors and Actuators B - Chemical }} \\
\emph{\textbf{Analytical Chemistry }}&\vline &\emph{\textbf{Talanta}} \\
Electroanalytical Chemistry &\vline & \\
\hline
& Applied Chemistry & \\
\hline
ACS Combinatorial Science &\vline & \emph{\textbf{Journal of Agricultural and Food Chemistry}} \\
\emph{\textbf{Dyes and Pigments }} &\vline & Journal of Combinatorial Chemistry \\
\emph{\textbf{Food Chemistry}} &\vline & \emph{\textbf{Microporous and Mesoporous Materials}} \\
\emph{\textbf{Food Hydrocolloids}} &\vline & Molecular Diversity \\
\hline
& Inorganic Chemistry & \\
\hline
Advances in Inorganic Chemistry &\vline & \emph{\textbf{Journal of Inorganic Biochemistry }}\\
\emph{\textbf{Dalton Transactions}} &\vline & Journal of Solid State Chemistry \\
\emph{\textbf{European Journal of Inorganic Chemistry}} &\vline & Organometallics \\
\emph{\textbf{Journal of Biological Inorganic Chemistry}} &\vline & \\
\hline
& Multidisciplinary Chemistry & \\
\hline
\emph{\textbf{ACS Nano}} &\vline & Journal of Controlled Release \\
\emph{\textbf{Angewandte Chemie - International Edition}} &\vline &\emph{\textbf{Journal of the American Chemical Society}} \\
\emph{\textbf{Chemical Science }}&\vline &\emph{\textbf{Tetrahedron - Asymmetry}} \\
Energy and Environmental Science &\vline & \\
\hline
&Organic Chemistry&\\
\hline
Advanced Synthesis and Catalysis &\vline & \emph{\textbf{European Journal of Organic Chemistry}} \\
Bioconjugate Chemistry &\vline & \emph{\textbf{Journal of Organic Chemistry}} \\
\emph{\textbf{Biomacromolecules}} &\vline & \emph{\textbf{Organic and Biomolecular Chemistry}} \\
Current Organic Chemistry &\vline & \emph{\textbf{Organic Letters}} \\
\hline
&Physical Chemistry&\\
\hline
ACS Catalysis &\vline & \emph{\textbf{Faraday Discussions}} \\
\emph{\textbf{Advanced Energy Materials}} &\vline & \emph{\textbf{Journal of Catalysis}} \\
\emph{\textbf{Advanced Functional Materials}} &\vline & \emph{\textbf{Journal of Chemical Theory and Computation}} \\
Advanced Materials &\vline & \emph{\textbf{Journal of Physical Chemistry B}} \\
Advances in Colloid and Interface Science &\vline & \emph{\textbf{Journal of Physical Chemistry C}} \\
Catalysis Science and Technology &\vline & \emph{\textbf{Journal of Physical Chemistry Letters}} \\
\emph{\textbf{ChemCatChem}} &\vline & \emph{\textbf{Langmuir}} \\
\emph{\textbf{Chemistry of Materials }}&\vline & \emph{\textbf{Physical Chemistry Chemical Physics}} \\
\emph{\textbf{Colloids and Surfaces B - Biointerfaces }}&\vline & Structure and Bonding \\
\hline
& Polymer Science & \\
\hline
Advances in Polymer Science &\vline & Macromolecular Rapid Communications \\
\emph{\textbf{Carbohydrate Polymers }}&\vline & \emph{\textbf{Macromolecules}} \\
\emph{\textbf{Journal of Membrane Science}} &\vline &Plasma Processes and Polymers \\
Journal of Polymer Science A - Polymer Chemistry &\vline &\emph{\textbf{Polymer Chemistry UK}} \\
\emph{\textbf{Macromolecular Bioscience}} &\vline &\emph{\textbf{Soft Matter}} \\
\hline
\end{tabular}%
\caption{The 67 chemistry journals listed in this Table were used,
for 2013, in our study.  Cross-disciplinary citations between
chemistry, molecular biology, and physics, from Brazil, South Korea,
Turkey, and USA, were searched from the 46 journals emphasized by
bold italics, as described in Sec.II.} \label{tab:addlabel}
\end{table*}

\begin{table*}[h!]
\begin{tabular}{! \vrule c c c ! \vrule }
\hline
&General Molecular Biology&\\
\hline
Biochimica Biophysica Acta: Molecular Cell Research &\vline & Molecular and Cellular Biology \\
Cell &\vline & \emph{\textbf{Molecular and Cellular Proteomics}} \\
\emph{\textbf{Journal of Molecular Biology }}&\vline & Molecular Biology of the Cell \\
Molecular Microbiology &\vline & Molecular Plant \\
Molecular Cell &\vline & \emph{\textbf{Oncogene}} \\
Molecular Biology and Evolution &\vline & \emph{\textbf{PLoS Computational Biology}} \\
Molecular Aspects of Medicine &\vline & \emph{\textbf{PLoS Genetics}} \\
EMBO Journal &\vline & Structure \\
EMBO Reports &\vline & \\
\hline
& Biochemistry and Biophysics & \\
\hline
ACS Chemical Biology &\vline & \emph{\textbf{Journal of Biological Chemistry}} \\
Acta Crystallographica D: Biological Crystallography &\vline & J. Proteins: Structure, Function, Genetics \\
\emph{\textbf{Biochemical Journal}} &\vline & Journal of Structural Biology \\
Biophysical Journal &\vline & Nature Chemical Biology \\
FASEB Journal &\vline & Nature Structural and Molecular Biology \\
Journal of Applied Crystallography &\vline & New Phytologist \\
\hline
& Biotechnology and Biomaterials & \\
\hline
\emph{\textbf{Bio Materials}} &\vline &Nature Biotechnology \\
Biotechnology Advances &\vline &Nature Methods \\
\hline
\end{tabular}%
\caption{The 33 molecular biology journals listed in this Table were
used, for 2013, in our study.  Cross-disciplinary citations between
chemistry, molecular biology, and physics, from Brazil, South Korea,
Turkey, and USA, were searched from the 8 journals emphasized by
bold italics, as described in Sec.II.} \label{tab:addlabel}
\end{table*}

\begin{table*}[h!]
\centering
\begin{tabular}{! \vrule c c c ! \vrule }
\hline
Journal &\vline & Topic \\
\hline
\emph{\textbf{European Physical Journal A}} &\vline & Hadrons and Nuclei  \\
\emph{\textbf{European Physical Journal B}} &\vline & Condensed Matter and Complex Systems \\
\emph{\textbf{European Physical Journal C}} &\vline & Particles and Fields \\
European Physical Journal D &\vline & Atomic, Molecular, Optical and Plasma Physics \\
European Physical Journal E &\vline & Soft Matter and Biological Physics \\
European Physical Journal H &\vline & Historical Perspectives on Contemporary Physics \\
\emph{\textbf{European Physical Journal AP}} &\vline & Applied Physics \\
\emph{\textbf{European Physical Journal ST }}&\vline & Special Topics \\
\emph{\textbf{European Physical Journal PLUS}} &\vline & Archiving and Documentation \\
Europhysics Letters &\vline & General Interest Impact \\
\emph{\textbf{Physica A}} &\vline & Statistical Mechanics and its Applications \\
\emph{\textbf{Physica B }}&\vline & Condensed Matter \\
\emph{\textbf{Physica C }}&\vline & Superconductivity and its Applications \\
\emph{\textbf{Physica D}} &\vline & Nonlinear Phenomena \\
\emph{\textbf{Physica E}} &\vline & Low-dimensional Systems and Nanostructures \\
\emph{\textbf{Physical Review A}} &\vline & Atomic, Molecular, and Optical Physics \\
\emph{\textbf{Physical Review B }}&\vline & Condensed Matter and Materials Physics \\
\emph{\textbf{Physical Review C}} &\vline & Nuclear Physics \\
\emph{\textbf{Physical Review D}} &\vline & Particles, Fields, Gravitation, and Cosmology \\
\emph{\textbf{Physical Review E}} &\vline & Statistical, Nonlinear, and Soft Matter Physics \\
Physical Review X &\vline & Cross-Topic, Cross-Field, Cross-Disciplinary \\
\emph{\textbf{Physical Review Letters}} &\vline & General Interest Impact \\
\hline
\end{tabular}%
\caption{The 22 physics journals listed in this Table were used, for
2013, in our study.  Cross-disciplinary citations between chemistry,
molecular biology, and physics, from Brazil, South Korea, Turkey,
and USA, were searched from the 17 journals emphasized by bold
italics, as described in Sec.II.} \label{tab:addlabel}
\end{table*}

\begin{table*}[h!]
    \begin{tabular}{! \vrule ccccccccccccc ! \vrule}
    \hline
    Cross-Science (CS) &\vline & No. of Sci.&\vline & No. of CS &\vline & Ratio Sci. A &\vline &CS Cit. &\vline & CS Cit. &\vline & No. of CS\\
    Citation from &\vline & A Papers &\vline & Citing Sci. &\vline &CS Citing/ &\vline & per Sci. &\vline & Standard &\vline & Cited Sci. \\
    Science A to Science B &\vline & Considered &\vline & A Papers &\vline &Considered &\vline & A Paper &\vline & Deviation &\vline & B Papers\\

    \hline
    &&&&Brazil & & & & & & & & \\
    \hline
Chemistry to M. Biology &\vline &272 &\vline &144 &\vline &0.5294 &\vline & 2.4743 &\vline & 5.1083 &\vline &673 \\
Chemistry to Physics &\vline &272 &\vline &144 &\vline &0.5294 &\vline & 3.9485 &\vline & 7.1090 &\vline &1074 \\
M. Biology to Chemistry &\vline &11 &\vline &3 &\vline &0.2727 &\vline & 0.8182 &\vline & 1.4025 &\vline &9 \\
M. Biology to Physics &\vline &11 &\vline &1 &\vline &0.0909 &\vline & 0.0909 &\vline & 0.2875 &\vline &1 \\
Physics to Chemistry &\vline &140 &\vline &30 &\vline &0.2143 &\vline & 0.6214 &\vline & 2.3612 &\vline &87 \\
Physics to M. Biology &\vline &140 &\vline &3 &\vline &0.0214 &\vline & 0.0500 &\vline & 0.3841 &\vline &7 \\

    \hline
    &&&&South Korea & & & & & & & & \\
    \hline
Chemistry to M. Biology &\vline &295 &\vline &138 &\vline &0.4678 &\vline & 1.8712 &\vline & 3.2443 &\vline &552 \\
Chemistry to Physics &\vline &295 &\vline &179 &\vline &0.6068 &\vline & 5.0373 &\vline & 7.7048 &\vline &1486 \\
M. Biology to Chemistry &\vline &15 &\vline &6 &\vline &0.4000 &\vline & 1.3333 &\vline & 3.6998 &\vline &20 \\
M. Biology to Physics &\vline &15 &\vline &4 &\vline &0.2667 &\vline & 1.2000 &\vline & 2.6128 &\vline &18 \\
Physics to Chemistry &\vline &126 &\vline &27 &\vline &0.2143 &\vline & 1.3095 &\vline & 3.3129 &\vline &165 \\
Physics to M. Biology &\vline &126 &\vline &4 &\vline &0.0317 &\vline & 0.1746 &\vline & 1.3515 &\vline &22 \\
\hline
    &&&&Turkey &&& & & & & & \\
    \hline
Chemistry to M. Biology &\vline &293 &\vline &152 &\vline &0.5188 &\vline & 1.7986 &\vline & 2.9229 &\vline &527 \\
Chemistry to Physics &\vline &293 &\vline &146 &\vline &0.4983 &\vline & 4.1809 &\vline & 8.1105 &\vline &1225 \\
M. Biology to Chemistry &\vline &12 &\vline &7 &\vline &0.5833 &\vline & 3.0000 &\vline & 3.5355 &\vline &36 \\
M. Biology to Physics &\vline &12 &\vline &4 &\vline &0.3333 &\vline & 1.4167 &\vline & 2.4650 &\vline &17 \\
Physics to Chemistry &\vline &158 &\vline &55 &\vline &0.3481 &\vline & 1.2278 &\vline & 2.5256 &\vline &194 \\
Physics to M. Biology &\vline &158 &\vline &1 &\vline &0.0063 &\vline & 0.2468 &\vline & 3.0928 &\vline &39 \\
\hline
    &&&&USA & & & & & & & &\\
    \hline
Chemistry to M. Biology &\vline &307 &\vline &158 &\vline &0.5147 &\vline & 2.6808 &\vline & 4.8531 &\vline &823 \\
Chemistry to Physics &\vline &307 &\vline &174 &\vline &0.5668 &\vline & 4.3518 &\vline & 8.4530 &\vline &1336 \\
M. Biology to Chemistry &\vline &18 &\vline &7 &\vline &0.3889 &\vline & 1.6667 &\vline & 2.5197 &\vline &30 \\
M. Biology to Physics &\vline &18 &\vline &4 &\vline &0.2222 &\vline &  0.3889 &\vline & 0.8085 &\vline &7 \\
Physics to Chemistry &\vline &168 &\vline &44 &\vline &0.2619 &\vline & 1.4048 &\vline & 3.4645 &\vline &236  \\
Physics to M. Biology &\vline &168 &\vline &4 &\vline &0.0238 &\vline & 0.0357 &\vline & 0.2413 &\vline &6 \\
\hline
    \end{tabular}%
    \caption{Cross-science citations between chemistry, molecular biology, and physics, grouped by country.} \label{tab:addlabel}
\end{table*}

\begin{table*}[h!]
    \begin{tabular}{! \vrule ccccccccccccc ! \vrule}
    \hline
    Cross-Science (CS) &\vline & No. of Sci.&\vline & No. of CS &\vline & Ratio Sci. A &\vline &CS Cit. &\vline & CS Cit. &\vline & No. of CS\\
    Citation from &\vline & A Papers &\vline & Citing Sci. &\vline &CS Citing/ &\vline & per Sci. &\vline & Standard &\vline & Cited Sci. \\
    Science A to Science B &\vline & Considered &\vline & A Papers &\vline &Considered &\vline & A Paper &\vline & Deviation &\vline & B Papers\\
    \hline

&&Chemistry &&to M. Biology, &&M. Biology to &&Chemistry & & & & \\
    \hline
    Brazil &\vline &272, 11 &\vline &144, 3 &\vline &0.5294, 0.2727 &\vline &2.4743, 0.8182 &\vline &5.1083, 1.4025 &\vline &673, 9 \\
    South Korea &\vline &295, 15 &\vline &138, 6 &\vline &0.4678, 0.4000 &\vline &1.8712, 1.3333 &\vline &3.2443, 3.6998 &\vline &552, 20 \\
    Turkey &\vline &293, 12 &\vline &152, 7 &\vline &0.5188, 0.5833 &\vline &1.7986, 3.0000 &\vline &2.9229, 3.5355 &\vline &527, 36 \\
    USA &\vline &307,18 &\vline &158, 7 &\vline &0.5147, 0.3889 &\vline &2.6808, 1.6667 &\vline &4.8531, 2.5197 &\vline &823, 30 \\
    \hline

&&M. Biology &&to Physics, &&Physics to &&M. Biology & & & &\\
    \hline
    Brazil &\vline &11, 140 &\vline &1, 3 &\vline &0.0909, 0.0214 &\vline & 0.0909, 0.0500 &\vline & 0.2875, 0.3841 &\vline &1, 7 \\
    South Korea &\vline &15, 126 &\vline &4, 4 &\vline &0.2667, 0.0317 &\vline & 1.2000, 0.1746 &\vline & 2.6128, 1.3515 &\vline &18, 22 \\
    Turkey &\vline &12, 158 &\vline &4, 1 &\vline &0.3333, 0.0063 &\vline & 1.4167, 0.2468 &\vline & 2.4650, 3.0928 &\vline &17, 39  \\
    USA &\vline &18, 168 &\vline &4, 4 &\vline &0.2222, 0.0238 &\vline & 0.3889, 0.0357 &\vline & 0.8085, 0.2413 &\vline &7, 6 \\
    \hline

&&Physics &&to Chemistry, &&Chemistry to &&Physics & & & & \\
    \hline
    Brazil &\vline &140, 272 &\vline &30, 144 &\vline &0.2143, 0.5294 &\vline &0.6214, 3.9485 &\vline &2.3612, 7.1090 &\vline &87, 1074 \\
    South Korea &\vline &126, 295 &\vline &27, 179 &\vline &0.2143, 0.6068 &\vline &1.3095, 5.0373 &\vline &3.3129, 7.7048 &\vline &165, 1486 \\
    Turkey &\vline &158, 293 &\vline &55, 146 &\vline &0.3481, 0.4983 &\vline &1.2278, 4.1809 &\vline &2.5256, 8.1105 &\vline &194, 1225 \\
    USA &\vline &168, 307 &\vline &44, 174 &\vline &0.2619, 0.5668 &\vline &1.4048, 4.3518 &\vline &3.4645, 8.4530 &\vline &236, 1336 \\
    \hline

    \end{tabular}%
    \caption{Cross-science citations from Brazil, South Korea, Turkey, and USA, grouped by sciences.} \label{tab:addlabel}
\end{table*}

\begin{table}[h!]
    \begin{tabular}{! \vrule ccccccccc ! \vrule}
    \hline
    Cross-Science &\vline & Brazil &\vline & ~South~ &\vline & Turkey &\vline & ~~USA~~ \\
    Citation Ratios &\vline & &\vline & Korea &\vline & &\vline & \\
    \hline
    Chemistry &\vline &0.9265 &\vline &0.8746 &\vline &0.7816 &\vline &0.7134 \\
M. Biology &\vline &0.3636 &\vline &0.4667 &\vline &0.5833 &\vline &0.4444 \\
    Physics &\vline &0.2143 &\vline &0.2302 &\vline &0.3481 &\vline &0.2679 \\

    \hline

    \end{tabular}%
\caption{Fraction of publications giving cross-science citations
from chemistry (to molecular biology and/or physics), from molecular
biology (to physics and/or chemistry), and from physics (to
chemistry and/or molecular biology).} \label{tab:addlabel}
\end{table}


\begin{references}
\bibitem{RosvallBergstrom} Maps of random walks on complex networks reveal community structure, M. Rosvall and C. T.
Bergstrom, Proc. Nat. Acad. Sci. {\bf 105} (4) 1118-1123 (2008).
\bibitem{RosvallBergstrom2} Exploring the astronomy literature landscape, E. A. Henneken, A. Accomazzi, M. J. Kurtz, C. S. Grant, D. Thompson, E. Bohlen, S. S. Murray, M. Rosvall, and C. Bergstrom, in Astronomical Data Analysis
Software and Systems XVIII, Eds. D. A. Bohlender, D. Durand, and P.
Dowler, {\bf 411}, 384-387 (2009).
\bibitem{RosvallBergstrom3} The transmission sense of information, C. T. Bergstrom and R.
Rosvall, Biol. Philos. {\bf 26}, 159–176 (2011).
\bibitem{Usdiken} Centres and Peripheries: Research Styles and Publication Patterns in 'Top' US Journals and their European Alternatives, 1960-2010, B. \"{U}sdiken, J. Management Studies {\bf 51}, 764-789
(2014).
\bibitem{Schott1} International influence in science: beyond center and periphery, T. Schott, Social Science Research {\bf 17}, 219–38 (1988).
\bibitem{Schott2} Ties between center and periphery in the scientific world-system: accumulation of rewards, dominance and self-reliance in the center,
T. Schott, J. World-Systems Research {\bf 4}, 112–144 (1988).
\bibitem{Hwang} International Collaboration in Multilayered Center-Periphery in the Globalization of Science and Technology, K. Hwang, Science
Technology Human Values {\bf 33}, 101-133 (2008).

\end{references}
\end{document}